\documentclass[12pt]{iopart}
\usepackage{graphicx}

\usepackage{bm}%
\usepackage[T1]{fontenc}
\expandafter\ifx\csname package@font\endcsname\relax\else
 \expandafter\expandafter
 \expandafter\usepackage
 \expandafter\expandafter
 \expandafter{\csname package@font\endcsname}%
\fi
\ifx\pdfoutput\undefined
\usepackage{graphicx}
\else
\fi
\usepackage{hyperref}

\hypersetup{
    bookmarks=false,         
    unicode=false,          
    pdftoolbar=false,        
    pdfmenubar=false,        
    pdffitwindow=false,     
    pdfstartview={FitH},    
    pdfnewwindow=true,      
    colorlinks=true,       
    linkcolor=black,          
    citecolor=blue,        
    filecolor=blue,      
    urlcolor=blue           
}

\usepackage{natbib}
\bibpunct{(}{)}{,}{s}{,}{,}

\begin{document}

\title[Effects of quasiparticle tunneling in a circuit-QED realization of a strongly driven...]{Effects of quasiparticle tunneling in a circuit-QED realization of a strongly driven two-level system}

\author{J Lepp\"akangas$^1$, S E de Graaf$^1$, A Adamyan$^1$, M Fogelstr\"om$^1$, A V Danilov$^1$, T Lindstr\"om$^2$, S E Kubatkin$^1$ and G Johansson$^1$}

\address{$^1$ Department of Microtechnology and Nanoscience, MC2, Chalmers University of Technology, SE-41296 G\"oteborg, Sweden}
\address{$^2$ National Physical Laboratory, Hampton Road, Teddington, TW11 0LW, UK}
\ead{juha.leppakangas@chalmers.se}

\begin{abstract}
We experimentally and theoretically study the frequency shift of a driven cavity coupled to a superconducting charge qubit. In addition to previous studies, we here also consider drive strengths large enough to energetically allow for quasiparticle creation. Quasiparticle tunneling leads to the inclusion of more than two charge states in the dynamics. To explain the observed effects, we develop a master equation for the microwave dressed charge states, including quasiparticle tunneling. 
A bimodal behavior of the frequency shift as a function of gate voltage can be used for sensitive charge detection.
However, at weak drives
the charge sensitivity is significantly reduced by non-equilibrium quasiparticles, which induce transitions to a non-sensitive state.
Unexpectedly,
at high enough drives, quasiparticle tunneling enables a very fast relaxation channel to the sensitive state. In this regime, the charge sensitivity is thus robust against externally injected quasiparticles and the desired dynamics prevail over a broad range of temperatures.
We find very good agreement between theory and experiment over a wide range of drive strengths and temperatures.
\end{abstract}



\maketitle

\section{Introduction}
The combination of small Josephson junctions with superconducting microelectronics offer
a versatile playground for quantum physics and quantum information,
often called circuit-quantum electrodynamics or circuit-QED~\cite{GeneralCircuitQED}.
In particular, it is possible to engineer circuits that realize a number of fundamental quantum optical systems.
Due to remaining unwanted noise sources, it is often necessary to describe the circuit dynamics in the framework of open quantum systems,
where decoherence plays an important role~\cite{Decoherence2005}.
However, one can also use engineered dissipation to drive the system towards a desired non-trivial steady-state.\cite{BathEngineeringPRL1996}
Both these aspects of dissipation are important in the system presented in this article.

Here, we study experimentally and theoretically a well known quantum optics problem realized within circuit-QED,
a strongly driven two-level system~\cite{chris2}. Previously it has been understood,
that in addition to the properties of the formed dressed states,
measurable quantities can depend crucially on the dissipation induced by interaction with the environment.
In Refs.~\cite{chris2,chris1} this was studied in the case of
longitudinally coupled noise and, for example, a population inversion
induced by the dissipative environment was experimentally verified.
The system was studied also in Ref.~\cite{sill} as a circuit-QED realization of a
Landau-Zener-St\"uckelberg interferometer~\cite{review}, and was found to be very potential
for accurate charge sensing. In this article, we present experimental results and
develop theoretical means for a very strong driving regime,
large enough to energetically allow for quasiparticle creation,
leading to new type of dissipation mechanisms.

\begin{figure}[b!]
\begin{centering}
\includegraphics[height=5cm]{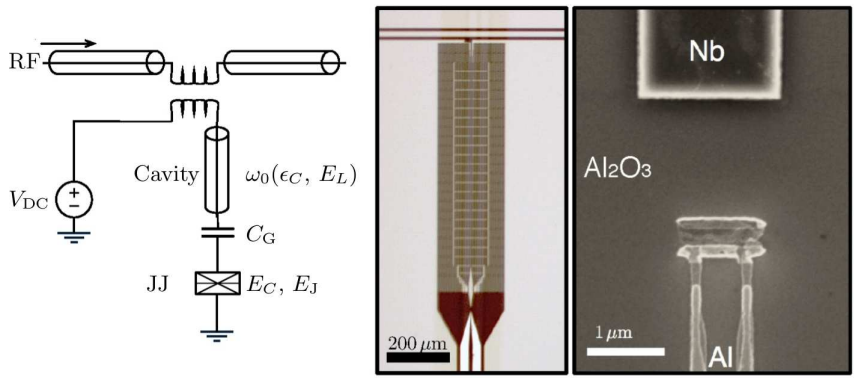} 
\caption{Left: Circuit representation of the sample. The gate charge $C_{\rm G}V_{\rm DC}$ is used
to control the number of Cooper pairs on the superconducting island between the
Josephson junction (JJ) and the gate capacitor $C_{\rm G}$. The shift in the resonance frequency
of the microwave cavity is probed by the inductively coupled transmission line.
Middle: Optical image of the sample. The unusual layout of the microwave cavity (grey area) is used to avoid Abrikosov vortices
and flux trapping.
Right: Scanning electron micrograph of
the Aluminium Cooper-pair box capacitively coupled to the Niobium microwave cavity
(corresponding to the black area in the optical image).}
\label{fig:sample}
\end{centering}
\end{figure}

The system is realized experimentally as a driven superconducting resonator (cavity)
capacitively coupled to a Cooper-pair box~\cite{Nakamura} (two-level system), see figure~\ref{fig:sample}.
The frequency shift of the cavity carries information of the formed dressed states and
is detected using a heterodyne null-detection technique called Pound-Drever-Hall locking, commonly used in optics and frequency metrology~\cite{black,tobias}.  We make the unexpected experimental observation, that in the quasiparticle dominated
regime the system has an increased sensitivity to changes in the gate charge of the Cooper-pair box~\cite{ShortPaper}.
This increase comes together with a change in the interference pattern,
compared to the previously studied regime at lower drive strengths.
This new pattern is stable up to relatively high temperatures and is robust towards externally produced nonequilibrium quasiparticles.

To explain the observed effects we add quasiparticle tunneling to the previously established density-matrix theory for the microwave dressed charge states.\cite{chris2} We consider both tunneling of existing quasiparticles as well as the creation of new quasiparticles at the Josephson junction. 
Quasiparticle tunneling leads to incoherent transitions between the even and odd electron-number parities of the superconducting charge qubit~\cite{LutchynFirst,Aumentado,Lenander,Sun}.
Here, the dressed charge-state dynamics in each separate parity state can be reduced to a two-level system. 

At weak drive strengths, the population of each two-level system is determined by the dissipation induced by the longitudinally coupled
charge-noise environment. At dressed-state resonances, there is population inversion in one of the parity subspaces. The corresponding bimodal behavior of the frequency shift with respect to the gate voltage can be used for sensitive detection of the gate charge.
This sensitivity is reduced by externally injected, or "nonequilibrium", quasiparticles, which induces incoherent transitions between the two parities, thus reducing the time spent in the parity state that gives high sensitivity.

At higher drive strengths, quasiparticle creation through photon-assisted tunneling is enabled.
This dissipative process dominates over the charge-noise environment and establishes a new pattern of population inversion.
Somewhat unexpectedly, the photon-assisted quasiparticle tunneling preferably drives the system towards the parity state with the highest charge sensitivity.  The dynamics is thus robust against externally injected quasiparticles and prevail for a wide range of temperatures.
The theory provides a detailed explanation of several new experimental phenomena seen in this high-drive regime.

This article is organized as follows. In Section~\ref{sec:CPB}, we introduce the established model of a superconducting charge qubit coupled to a cavity.
We briefly describe the master equation for the system density matrix in the basis of dressed qubit charge states,
and then extend it to account for quasiparticle creation at large drives.
In section~\ref{sec:Numerics}, we discuss the numerical methods used to solve the eigenstates and the steady-state density matrix,
and how we estimate the frequency shift of the cavity from this information.
In Section~\ref{sec:experiment}, we discuss the details of the experimental realization. The theory and experiment
are then compared in Section~\ref{sec:results}.
Specifically, we consider the frequency shift and its temperature dependence. Based on the good agreement between theoretical and experimental results,
we make a detailed discussion of the physical processes behind the frequency shift pattern. Conclusions are given in Section~\ref{sec:Discussion}.

\section{The system and the model}\label{sec:CPB}

In this section, we introduce our model of the superconducting charge qubit coupled to the driven microwave cavity.
After establishing the Hamiltonian, we consider an optimal basis for its numerical diagonalization, i.e. the displaced number states.
We then consider the effect of the external probe, which drives the cavity state towards a large amplitude coherent state, allowing for a semiclassical approximation in the photon basis.

We continue by introducing the well known Bloch-Redfield master equation, which is used to model the generalized populations of the dressed charge states in the presence of external charge fluctuations.
After this, we straightforwardly extend this approach to account for quasiparticle tunneling. We discuss in detail
two energetically very different processes, i.e.~tunneling of existing quasiparticles and tunneling with quasiparticle creation.
At the end of this section, we discuss how to calculate measurable properties in the system within this approach,
such as the frequency shift of the cavity and the energy absorption rate induced by quasiparticle creation, 
experimentally measurable from the response of the system to the external drive.

\subsection{Hamiltonian and basis states}
A circuit representation of the experiment is shown in figure~\ref{fig:sample}.
The system consists of a Cooper-pair box coupled through a capacitor $C_{\rm G}$ to a microwave cavity of angular frequency $\omega_0$.
The Hamiltonian of the system has the general form
\begin{eqnarray}
H&=&\hbar\omega_0 a^{\dagger}a+E_C(\hat n-n_{\rm G})^2\nonumber \\
&&-\frac{E_J}{2} \sum_{n}\left( \vert n+2 \rangle\langle n \vert +\vert n\rangle\langle n+2 \vert  \right)+g(a^{\dagger}+a) \hat n.\label{eq:hamiltonian1}
\end{eqnarray}
Here $a^{(\dagger)}$ is a cavity photon annihilation (creation) operator and $\hat n$ is the island-charge-number operator, counted as electron charges and restricted to integer values. Here we include arbitrary number of charge states, but most of the observed effects can be understood
in a two-level approximation, discussed below.
The term proportional to $E_{C}=e^2/(C+C_{\rm G})$
corresponds to the Coulomb energy of the island,
that can be affected by
the dimensionless gate charge $n_{\rm G}=C_{\rm G}V_{\rm DC}/e$, controllable by the applied DC-voltage $V_{\rm DC}$.
Cooper-pair tunneling across the Josephson junction
appears as a coupling term proportional to the Josephson coupling energy $E_{\rm J}$,
coherently switching the electron number of the island by two.
The coupling between the Cooper-pair box and
the cavity is described by the last term on the right-hand side of~(\ref{eq:hamiltonian1}).
In circuit-QED this can be derived using a lumped-element model~\cite{juha1} which gives
$g=\hbar\omega_0G C_{\rm G} /C_{\rm J}$, where $G=(2\epsilon_C/E_L)^{1/4}$.
Here $G$ compares the oscillator's charging energy $\epsilon_C$ and its magnetic energy $E_L$.
For usual $\lambda/4$ transmission-line resonators this can be represented as
as the ratio between the wave impedance $Z_0$ and the resistance quantum $R_{\rm Q}=h/4e^2$, $G=(\pi Z_0/R_Q)^{1/2}$.
Usually in circuit-QED $G\sim 0.1-0.2$, but also stronger
coupling regimes have been proposed  by using Josephson metamaterials~\cite{Metamaterials,Metamaterials2}.
In the experiment considered in this article $G=0.12$ ($Z_0=30$~$\Omega$).

A convenient basis for further analysis is formed by the displaced number states,
\begin{equation}\label{eq:basis1}
\vert n;N\rangle^{0} = \exp[- \beta \hat n( a^{\dagger}- a)]\vert n\rangle\vert N\rangle = U(\beta,n)\vert n\rangle\vert N\rangle .
\end{equation}
Here, $N$ corresponds to the cavity photon number and
$D(\beta)$ is the displacement operator.
These are eigenstates of Hamiltonian (\ref{eq:hamiltonian1}) for $E_{\rm J}=0$ and $\beta=g/\hbar\omega_0$,
with the corresponding eigenenergies
\begin{equation}\label{eq:eigenenergies1}
E_{nN}^0=N\hbar\omega_0 +E_C(n-n_G)^2-\hbar\omega_0 \left( \frac{C_{\rm G}}{C_{\rm J}} \right)^2G^2 n^2.
\end{equation}
The last term on the right-hand side is
very small and will be neglected in the following.

For finite $E_{\rm J}$, Cooper-pair tunneling coherently couples the charge states differing by two electron charges, i.e.~one Cooper-pair, as 
described by the third term on the right-hand-side of (\ref{eq:hamiltonian1}).
How the corresponding displaced photon states couple depends on their overlap. 
In the considered semiclassical limit, $N\gg 1$, the corresponding matrix elements are
\begin{equation}\label{eq:CouplingAmplitudes}
\langle n\pm 2;N+l\vert^0 \sigma_x \vert n ;N\rangle^0 \approx  J_{l}\left( \pm 4\beta\sqrt{N}\right)
\end{equation}
This lets us construct the Hamiltonian (\ref{eq:hamiltonian1}) in the dressed-states basis (\ref{eq:basis1})
and in the limit of high photon numbers.
We will label these states analogously as $\vert n;N\rangle$, since at high photon numbers $N$, the hybridized states form a
set of ladders (${\rm Dim}\{ n \}$ ladders), each
similar to harmonic oscillator states, but for $\beta\ll 1$ slightly changing energy intervals between the different steps
(values of $N$).
Here, we also note a useful connection between the Bessel function arguments and the classical gate oscillation
amplitude, $n_{\rm G}(t)=A\cos\omega_{\rm d} t$, as $\beta\sqrt{\langle N \rangle}= E_C A/\hbar\omega_0$, where $\langle N \rangle$
is the mean number of photons in the cavity.

From ~(\ref{eq:eigenenergies1}), we see that each time
\begin{equation}\label{eq:ChargingEnergy}
\delta E_C= E_C(n+ 2-n_{\rm G})^2 - E_C(n-n_{\rm G})^2 =l\hbar\omega_0 ,
\end{equation}
where $l$ is an integer, the eigenstates (for $E_{\rm J}=0$) are degenerate $E_{(n+2)(N-l)}=E_{n N}$. Close to these
dressed-state resonances (from here on we denote them as $l$-photon resonances) $E_{\rm J}$ opens up a gap between the degenerate dressed states with magnitude $\sim E_{\rm J}\vert J_{l}(4\beta\sqrt{N})\vert$.
For increasing $|l|$, an increasing drive is needed to open up the gap, which can be seen from the slow onset of the higher order Bessel functions [$J_l(x)\propto x^{|l|}, x\ll 1$].

Our Hamiltonian~(\ref{eq:hamiltonian1}) includes Cooper-pair tunneling, but not quasiparticle tunneling. This implies that
only states differing by two electron charges are coherently connected. Thus, we naturally define the even (odd) parity Hilbert space, consisting of the even (odd) electron number eigenstates $\vert n \rangle$.  Our Hamiltonian has a block-diagonal form with respect to the two charge parities.

\subsection{The effect of the external drive}\label{sec:eigenstates}
In this article, we treat the effect of the drive
and the corresponding dissipation due to the connection to the open
transmission line semiclassically.
This means that the drive enters the calculation as a finite value for the cavity amplitude $A$.
For the uncoupled cavity $g=0$, we know that the exact solution,
of a driven damped harmonic oscillator,
is a coherent state $\vert \alpha \rangle$~\cite{QOBook}.
For finite coupling $g$, the charge and the
photon degrees of freedom hybridize. However,
for weak coupling $\beta\ll 1$ and large average photon number $N\gg1$, 
we assume that the photon number distribution of the cavity is still strongly peaked around $N$ with a variance not much larger than $\sqrt{N}$.
For consistency, we also calculate the photon loss rate induced by the charge qubit and find that it is indeed small compared to the photon loss through the transmission line.

Further, we will approximate the amplitude dependent transition rates 
derived below as constant within the variance of the photon number distribution. 
This is indeed a good approximation in the weak coupling regime $\beta\ll 1$.
The validity of these approximations is further motivated by the good agreement between the theoretical and experimental results, discussed in section~{\ref{sec:results}}.

\subsection{Bloch-Redfield master equation including charge fluctuations}\label{sec:environment}
In addition to the coherent physics described in Section~\ref{sec:CPB},
the qubit-cavity system will also be influenced by decoherence
due to interaction with an open dissipative environment~\cite{Decoherence2005,Weiss}. For a superconducting charge qubit this includes
 gate-charge fluctuations,~\cite{Decoherence2005,Astafiev} considered here. In section~\ref{sec:QPTunneling}, we also include the tunneling and creation of quasiparticles at the Josephson junction.
Here we use a previously established Bloch-Redfield master equation~\cite{Bloch,Redfield} for the density matrix of the system, in the dressed charge states basis~\cite{chris2},
\begin{equation}\label{eq:densitymatrixreduced}
\dot\rho(t)= L_0(A)\rho(t) + \hat\Sigma_{\rm CF}(A) \rho(t) + \hat\Sigma_{\rm QP}(A) \rho(t),
\end{equation}
where we consider a small number of charge states and a finite number of photon states around the average photon number $N$, set by the drive amplitude.
The Liouville operator $L_0=(i/\hbar)[\cdot ,H]$ describes the coherent qubit-cavity interaction and
$\hat \Sigma_{\rm CF}$ is the generalized transition rate describing the effect of charge fluctuations, and $\hat \Sigma_{\rm QP}$ describes quasiparticle tunneling, discussed in section~\ref{sec:QPTunneling}.

The general transition rates in this approach have the form,
\begin{eqnarray}\label{eq:masterequation}
& &\Sigma_{b\rightarrow  n}^{a\rightarrow  m}=i\frac{E_b-E_a}{\hbar}\delta_{am}\delta_{bn} +\frac{C_{ma}C_{nb}^*}{2}\left[  \Gamma (E_{a}-E_{m})+ \Gamma(E_{b}-E_{n}) \right]  \nonumber \\
&-&\frac{1}{2}\sum_v\left[ C_{va}C_{mv}^*  \Gamma(E_{a}-E_{v})\delta_{bn} +  C_{nv}C_{vb}^*  \Gamma(E_{b}-E_{v})\delta_{am} \right].
\end{eqnarray}
Here $C_{ij}=\langle i\vert \hat C\vert j\rangle$, where $\hat C$ is the coupling operator discussed below,
and $\Gamma(E)$ is the energy dependent transition rate proportional to the corresponding spectral density of the environment.
For charge fluctuations, the coupling operator $\hat C=\hat n$ and to the energy dependent rates are
\begin{eqnarray}
\Gamma^{\rm CF}(E)=\frac{\pi \alpha E}{\hbar[1-\exp(-\beta E)]}.
\end{eqnarray}

We here consider an Ohmic environment with a dimensionless coupling parameter $\alpha$.
For a two-level system this is equivalent to the spin-boson model.~\cite{leggett}
In the experiment, we observe effects characteristic for this type of environment at low drive strengths, see Section~\ref{sec:results}.
In this approach, the coupling to the environmental degrees of freedom is assumed to be weak, and
higher-order processes can be neglected, which can be made for $\alpha\ll 1$.

In~(\ref{eq:masterequation}), for simplicity, the generalized transmission rate is also assumed to be real valued.
The real parts give dissipative transitions between the basis states, while the imaginary parts gives shifts of the energy levels.
The imaginary parts are of the same size as the real parts, which are small compared to the typical energy-level splittings in our system.

\subsection{Effect of quasiparticle tunneling}\label{sec:QPTunneling}
The main subject of this article is the effect of
quasiparticle creation occuring at large drive strengths.
We will here derive the corresponding amplitude-dependent transition rates $ \hat\Sigma_{\rm QP}(A) $ to include in the master equation (\ref{eq:densitymatrixreduced}).
The main reason we can use such a perturbative description, is because we are considering a tunnel junction, i.e.~the tunnel resistance $R_{\rm T}$ is larger than the resistance quantum $R_{\rm Q}$, and that typical energy differences are much lower than the superconducting gap, $\Delta$.
Thus, the main effect of the quasiparticle tunneling is to introduce incoherent transitions between the two electron-number parity eigenstates~\cite{LutchynFirst,Aumentado,Lenander,Sun}, which have so far been uncoupled.
It is important to notice, that we do not model the macroscopic number of electronic
degrees of freedom in this approach. They enter perturbatively through the
quasiparticle density of states and distributions, as discussed below.
In the master equation, we only keep track of changes in the total charge on the island.

\subsubsection{Energy spectrum of quasiparticle processes}
We include quasiparticles perturbatively
in the Bloch-Redfield approximation.
This means that quasiparticle states appear
through their density of states, their corresponding populations, and their tunneling amplitudes across the Josephson junction.
As the tunneling amplitudes are considered to be energy independent, the relevant information comes from the first two properties.

\begin{figure}[b!]
\begin{centering}
\includegraphics[width=0.9\linewidth]{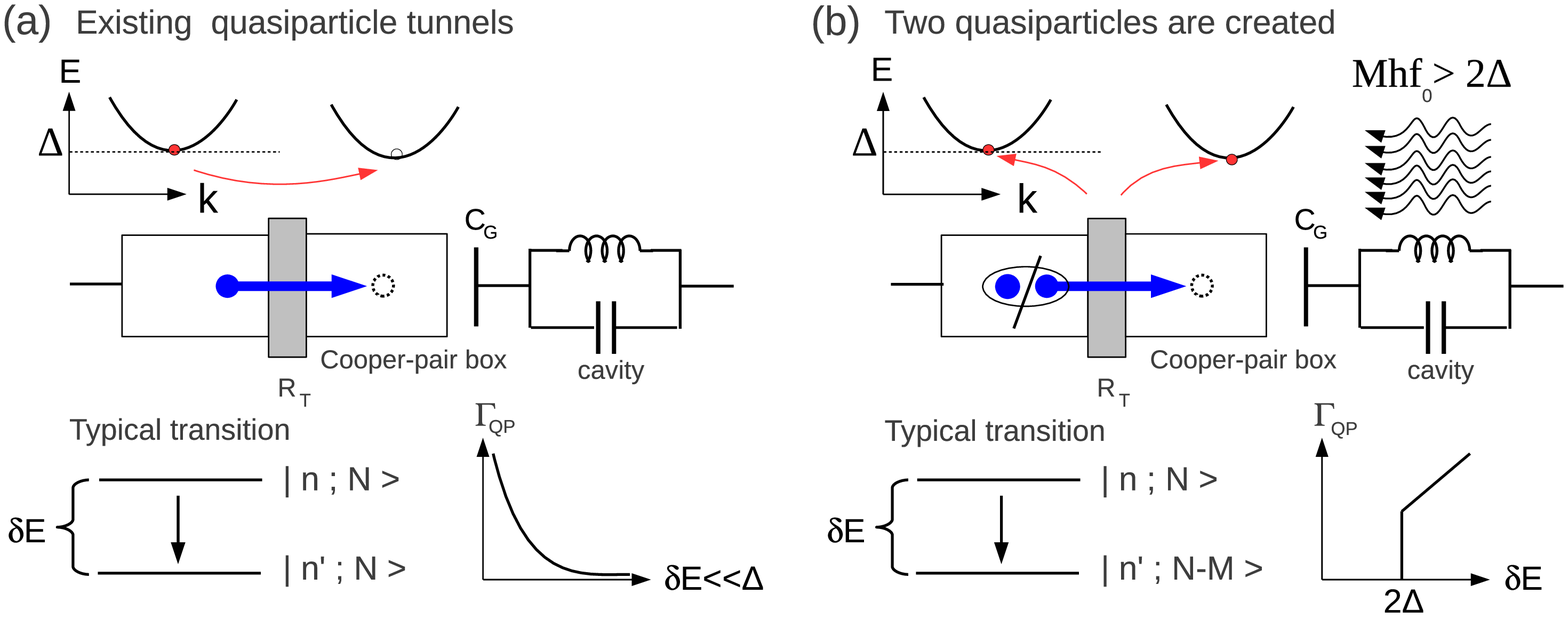}
\caption{Visualization of the two relevant quasiparticle
processes. Electron-tunneling across the Josephson junction occur through (a) tunneling of an existing quasiparticle excitation (thermal or nonequilibirum) or (b) through the creation of two quasiparticles, one on each side of the junction.
Process (a) does not change the total number of quasiparticle
excitations and the energy-level difference $\delta E$ goes to/is absorbed from the quasiparticle's kinetic degrees of freedom.
This process occurs typically together with a transition between eigenstates having a small energy difference, as the
corresponding rate $\Gamma_{\rm QP}$ increases toward zero energy difference.
Process (b) creates two new quasiparticles and the rate has a threshold-like behaviour at $2\Delta$.
Eigenstates separated by more than $M$ photons, where $M\hbar\omega_0 \geq 2\Delta$,
become well coupled at large drive strengths. ($M\approx 14$ in the experiment.)}
\label{fig:QPVisualization}
\end{centering}
\end{figure}

The normalized (BCS) density of states has a very sharp energy dependence~\cite{Tinkham},
$\rho(\omega)=\Theta(\omega^2-\Delta^2)\vert \omega\vert/\sqrt{\omega^2-\Delta^2}$, where $\Theta(x)$ denotes a unit step function.
This gives rise to a different energy dependence of the quasiparticle rates, compared to the Ohmic density of states for charge fluctuations considered in section~\ref{sec:environment}.
The quasiparticles are fermions, and we consider their population to be given by the equilibrium Fermi distribution
$f(\omega)$.
The effective rate of all quasiparticle states, $\Gamma(E)$, is the same as the usual quasiparticle tunneling rate
between two superconductors~\cite{Tinkham}
\begin{equation}
\Gamma^{\rm QP}( E)=\frac{1}{R_{\rm T}e^2}\int_{-\infty}^{\infty}d\omega \int_{-\infty}^{\infty}d\omega'f(\omega)(1-f(\omega')) \rho(\omega)\rho(\omega') \delta(\omega-\omega'+ E).\label{eq:quasirates}
\end{equation}
The tunneling resistance of the Josephson junction $R_{\rm T}$ is
defined by the Ambegaokar-Baratoff formula $R_{\rm T}=R_{\rm Q}\Delta/2E_{\rm J}$ ($T=0$).
This function has a very nonlinear form, diverging
at finite temperatures at $E=0$, and having
step-like threshold activation at $E=2\Delta$,
as visualized in figure~\ref{fig:QPVisualization}.

Quasiparticle tunneling processes occur physically through single-electron tunnelings across the Josephson junction. This is
included in the considered Bloch-Redfield master-equation by the two coupling operators~\cite{juha3},
\begin{equation}
\hat C^+=\sum_n\vert n+1\rangle\langle n\vert\, ,  \,\,\,\,\,\, \hat C^-=\sum_n\vert n-1\rangle\langle n\vert\ . \label{eq:quasioperators}
\end{equation}
For simplicity, we sum the two electron-tunneling directions ($\hat C$ and $C^{\dagger}$) incoherently (separately),
 since
based on our numerical simulations with a more general master equation, the possible interference between
electron and hole-like quasiparticle tunneling~\cite{Lutchyn,Martinis,Catelani,juha4} is not observed in the considered experiment.

\subsubsection{Nonequilibrium quasiparticle distributions}
It is known that in typical experimental realizations, the quasiparticle distribution can deviate significantly from the equilibrium Fermi distribution~\cite{Lutchyn}. 
E.g.~hot quasiparticles can diffuse through leads from higher temperature regions\cite{Martinis}. 
We use here a simplified model for such a nonequilibrium situation,
by considering an increased temperature of the quasiparticles, compared to the temperature of the dilution refrigerator.
This is important in the low drive strength region, where
nonequilibrium quasiparticle effects dominate, see section~\ref{sec:results}.
For large drives, where system itself creates quasiparticles, the dynamics becomes immune to changes in the quasiparticle density,
and thus no further changes in the effective quasiparticle temperature was needed to be introduced.

A special nonequilibrium situation occurs when the number of electrons on the island is odd.
In this case, all electrons cannot pair to form Cooper pairs, and there exist at least one quasiparticle excitation on the island.
The dynamics of this single excitation has been studied
in detail~\cite{Schon,LutchynFirst}. It causes an asymmetry between the transition rates,
dependent on from which parity state the system makes the transition.
This is the effect of the single extra quasiparticle
tunneling out of the island. Here, it is modeled in the simplest way, by adding an extra rate $\Gamma_{\rm odd}\Theta(E)$ to the total transition rate out of the odd parity states~\cite{juha3}.

\subsubsection{Quasiparticle processes}\label{sec:QPProcesses}
Quasiparticle tunneling, described by
expressions~(\ref{eq:quasirates}-\ref{eq:quasioperators}), can be divided into two energetically very different processes.
At low drive strengths, the dominating process is tunneling of quasiparticles already existing in the leads~\cite{Lutchyn,Martinis}.
The process does not change the number of quasiparticle excitations and the energy difference $\delta E$ goes to/is absorbed from the quasiparticle's kinetic degrees of freedom.
This process is decribed by the quasiparticle tunneling rate $\Gamma(E)$ for $E<2\Delta$.
Here, most of the quasiparticles have energies nearby the superconducting gap, where the BCS-density of states diverges. 
This leads to a divergence of the tunneling rates for zero energy difference~\cite{Tinkham}, see figure~\ref{fig:QPVisualization}a.

At large drive strengths, electron tunneling by creation of quasiparticles becomes possible. Here, a single Cooper-pair
breaks on one side of the Josephson junction. One of the unpaired electrons then tunnels and forms another
quasiparticle on the other side~\cite{juha5}, see figure~\ref{fig:QPVisualization}b.
This process creates two new quasiparticles and
its rate therefore has an energy threshold given by the superconducting energy gap $2\Delta$.
This energy can be extracted from transitions between eigenstates separated by more than $M$ photons, where $M\hbar\omega_0 \geq 2\Delta$.
The matrix elements for these single-electron tunneling transitions (in the relevant semiclassical limit $N\gg1$) are
\begin{eqnarray}\label{eq:amplitudeqp}
\langle n +(-) 1;N+M\vert \hat C^{(\dagger )} \vert n;N\rangle &\approx& J_{M}\left[ +(-) 2\beta\sqrt{N}\right].
\end{eqnarray}
These matrix elements are non-zero only at higher drive strengths, due to the slow onset of the $M$th order Bessel function, $J_M(x) \propto x^M, x\ll1$.
Physically, this corresponds to a process where $M$ photons are simultaneously absorbed from the oscillator to split a Cooper pair, as visualized in figure~\ref{fig:QPVisualization}b.

\section{Numerical methods}\label{sec:Numerics}
\subsection{Diagonalization and the reduced master equation}
We numerically solve for the eigenstates of Hamiltonian~(\ref{eq:hamiltonian1}) in a certain range [$N_{\rm m}-\delta N,N_{\rm m}+ \delta N$] around the mean photon value $\langle N\rangle=N_{\rm m}$, in the displaced photon-number basis~(\ref{eq:basis1}).
We use the semiclassical approximation for the off-diagonal coupling amplitudes~(\ref{eq:CouplingAmplitudes}), and because of the small coupling $\beta\ll1$, we can neglect their amplitude dependence over the photon number range $2\delta N$.
In the numerical results presented in this article (section~\ref{sec:results}),
we used six charge states, three for each parity, and $\delta N=35$ giving $70$ photon states symmetrically around the used mean photon value $N_{\rm m}\sim 10^3-10^5$
($A\sim 0.1-1$).
Even though the resulting width of the eigenstates
in photon space was much smaller, the high photon number was needed to minimize truncation errors.

The diagonalization gives $\rm{Dim}\{ n \}\times (2\delta N+1)$ eigenstates. Since we neglect the small amplitude dependence of the off-diagonal coupling elements in the Hamiltonian, the eigenenergies are grouped in equidistant harmonic oscillator ladders, up to truncation errors close to the end of the photon number range. The number of ladders equals the number of included charge states.
The first difficulty is to identify and sort the obtained eigenstates into this photon-ladder form.
To identify the central dressed cavity states, we pick up $\rm{Dim}\{ n \}$ states that minimize the expectation value $\langle \hat N-N_{\rm m} \rangle$, as
such states  cannot be translational equivalent. We label these as $\vert n ; N\rangle$ (important here is the variable $N$,
how to label the new charge states $n$ for the photon number $N$ is not important).

%

After identification of the eigenstates,
we construct the Bloch-Redfield master-equation depicted in sections \ref{sec:environment}-\ref{sec:QPTunneling},
to solve for the generalized populations. 
In this step, we trace out the photonic degrees of freedom and solve for the reduced density matrix of the dressed charge states
\begin{equation}\label{eq:densitymatrix3}
\tilde\rho_{nn'}\equiv \sum_{N} \langle n;N\vert \rho \vert n';N\rangle.
\end{equation}
The assumption of translation invariance of $L_0$ and $\hat\Sigma$ lets us to define the reduced transition-rate tensor
\begin{equation}\label{eq:densitymatrix5}
\tilde\Sigma_{n_{i'}\rightarrow n_{f'}}^{n_i\rightarrow n_f}(A)=\sum_{\Delta N}\hat\Sigma_{n_{i'}N_{\rm m}\rightarrow n_{f'}(N_{\rm m}+\Delta N)}^{n_iN_{\rm m}\rightarrow n_f(N_{\rm m}+\Delta N)}(A).
\end{equation}
Here the states $\vert n_i;N_{\rm m}\rangle$ and $\vert n_{i'};N_{\rm m}\rangle$ belong to the chosen group of central eigenstates,
from which the ${\rm Dim}\{ n \}$ photon ladders can be constructed.
The master equation for the dressed charge states has the same form as~(\ref{eq:densitymatrixreduced}),
but is written for the central dressed charge states $\vert n;N_{\rm m}\rangle$ using the reduced transition-rate tensors $\tilde \Sigma (A)$.

Nondiagonal entries of the density matrix within the same $N$ (states $\vert n ; N\rangle$) are included within this approach.
This accounts for situations where decoherence exceeds an energy-level difference between two dressed states,
leading to a Zeno effect~\cite{shnirman,juha3}, where environmental disturbance significantly inhibits coherent coupling between two nearly degenerate states.
It is observed also here, that including nondiagonal entries removes some very sharp spurious resonances, that are obtained when
including only populations (diagonal entries of $\tilde \rho$).

\subsection{Estimating measured quantities}\label{sec:FreqShift}
The photonic degrees of freedom are now traced out, and enter the master equation~(\ref{eq:densitymatrixreduced}) only through the drive amplitude $A$.
We will now deduce two measurable properties of the system, the frequency shift of the cavity and the power dissipated in the charge fluctuators and quasiparticle continuum.
The frequency shift, a very small fraction of $\omega_0$, is due to the hybridization of the charge and oscillator degrees of freedom.
To deduce this from the dressed charge-state simulation, we calculate the dressed charge states energy levels $E_{n}(A)$ as a function of the amplitude $A$, using small discrete steps $\delta A$.
For small $\beta$, the frequency-shift related to the dressed state $\vert n;N\rangle$ can then be estimated as
\begin{equation}
\delta f_0^{n}(A)=\frac{E_{n}(A+\delta A)-E_{n}(A)}{\delta A}\times \frac{\partial A}{\partial \langle N \rangle}.
\end{equation}
In the numerical simulations presented in
section~\ref{sec:results}, we used $\delta A = 3\times 10^{-3}$ corresponding approximately to shifts of $3\times 10^2$ photons.
Using the steady-state density matrix $\tilde\rho^{\rm s}$ for the dressed charge states, we estimate the measured frequency shift as
\begin{equation}\label{eq:FregShift2}
\langle \delta f_0(A) \rangle= \sum_{n }\tilde \rho_{nn}^{\rm s}(A)\delta f_0^{n}(A).
\end{equation}

We can also estimate the dissipation rate of photons due to the interaction with the charge fluctuators and quasiparticle environment. 
This quantity can be used as an indicator for the onset of different processes as the drive is increased. The photon loss rate can be
estimated as
\begin{equation}\label{eq:densitymatrix6}
\Gamma_{\rm Ph} =\sum_{n_i,n_{i'},n_f,\Delta N} \Delta N \times \hat\Sigma^{n_i N_{\rm m}\rightarrow (n_f N_{\rm m}-\Delta N)}_{n_{i'} N_{\rm m}\rightarrow (n_f N_{\rm m}-\Delta N)}\tilde\rho^{\rm s}_{ii'}(A).
\end{equation}
The dissipated power is then simply $P=\hbar\omega_0 \Gamma_{\rm Ph}$.

\section{Experimental realization}\label{sec:experiment}

Measurements were performed on a sample which layout is shown in figure~\ref{fig:sample}. The sample combines a niobium resonator on sapphire with an aluminum Cooper-pair box made with two-angle evaporation and oxidation. The cavity is conceptually equivalent to a magnetically coupled $\lambda/4$-resonator. Both to reduce its length and make it resilient to flux \cite{fractalpaper} it is loaded with an interdigitated capacitance to ground. This results in a somewhat reduced impedance (30~Ohms) and a lower propagation velocity. The magnetic coupling of the resonator ensures a decoupling of the
Cooper-pair box to other modes of the microwave field, and provides for a better isolation from the environment\cite{kim2012,houck2008}. In our device the resonator is close to critically coupled with a loaded quality factor of $4\times 10^4$. The coupling between cavity and Cooper-pair box was designed to be weak, considerably smaller than the junction capacitance. DC biasing of the box is obtained through the same leads as the microwave excitation of the Cooper-pair box, see figure~\ref{fig:sample}.

Since the frequency shift due to transitions between dressed states is expected to be very small we require a very high accuracy when determining the resonator frequency. A high Q cavity partly overcomes this issue since the relative phase change due to a small frequency deviation is directly proportional to the cavity line-width. However, we still need a very good frequency readout precision that is also relatively fast, not to suffer from slow drifts. Such drifts can for example be intrinsic to the cavity and related to two-level fluctuators and 1/f noise\cite{jonspaper}, charge jumps in the vicinity of the Cooper-pair box that shifts the island charge a fraction of an electron \cite{oneoverfnoise} or thermal fluctuations. For this reason we have used a technique called Pound-Drever-Hall (PDH) locking \cite{black,tobias}. Similar to heterodyne detection, this technique is based on the null detection of the phase interference between a carrier and a reference signal. In the PDH scheme both these signals are passed through the same cables in the form of a phase modulated (PM) spectrum. After probing the cavity with this PM spectrum we amplify the signal (using a low noise HEMT with $T_n\approx4$ K) and after careful filtering the spectrum is mixed in a diode detector at room temperature.
When the phase of the lower PM sideband is shifted an equal amount (but opposite sign) as the upper sideband relative to the carrier, (i.e. at resonance), their phase will exactly cancel out, and there will be no output signal from the diode at the sideband frequency. However, slightly off-resonance the phases will be different, $\vert\phi^{sb}_{-1}-\phi^{\rm{carrier}}\vert \neq \vert\phi^{sb}_{+1}-\phi^{\rm{carrier}}\vert$, the mixing in the diode will result in a beating phase and a signal at the frequency $\omega_{\rm{carrier}}-\omega_{-1}^{sb}$, i.e. an amplitude modulation at the sideband frequency. This signal can easily be detected with a lock-in amplifier. A PID controller is then given the task to null this beating signal by adjusting the carrier frequency (using a voltage-controlled oscillator). The output from the PID is thus directly proportional to the frequency shift of the resonator and can be used both to track and to measure the center frequency with very high precision and bandwidth. For a detailed explanation of our experimental setup, see Ref. \cite{degraaf2013}. The fact that the carrier and the reference is passed through the same cables effectively eliminates drifts due to thermal and mechanical fluctuations since this type of noise becomes correlated and cancels (to first order) in the heterodyne mixing. Furthermore, since the phase interference originates from the mixing of the two PM sidebands with the carrier, the sidebands can be placed well outside the resonator line-width not to interfere with the coherent excitation of the cavity.

In this experiment, the PID is not operating as a way to control a purely quantum mechanical system.
Rather, it tracks the average (classical) state of the cavity on a timescale much slower than the cavity frequency.
The coherent excitation is thus slowly steered to always be on resonance.

\begin{figure}
\begin{centering}
\includegraphics[width=\linewidth]{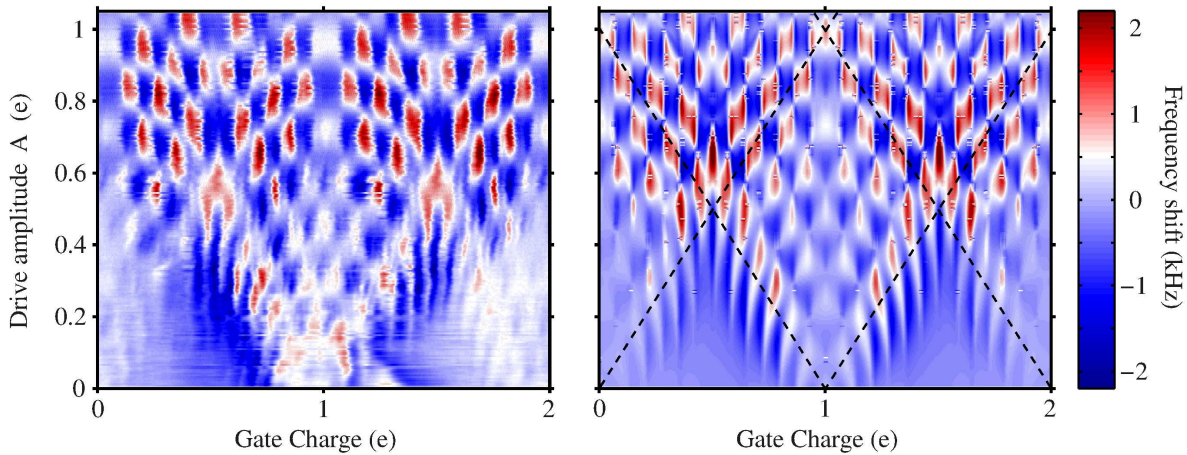}
\caption{Left: The measured frequency shift as a function of the gate charge and drive amplitude ($T=20$~mK).
Right: Numerical results based on the theoretical model described in sections~\ref{sec:CPB} and~\ref{sec:Numerics}.
We apply a Gaussian gate-charge averaging of width 0.0052e to the numerical results,
that accounts for low-frequency gate-charge fluctuations~\cite{TempDep1,TempDep2}.
The measurement is done at a relatively short timescale ($\sim$1s/gate trace),
and we thus suffer less from 1/f noise.
The parameters in the numerical simulation are $\omega_0/2\pi=6.94$~GHz, $E_{\rm J}/h= 4.82$~GHz, $E_C/h= 24.4$~GHz, $\Delta/h= 48.2$~GHz, $C_{\rm G}/C_{\rm J} = 1/18$, $T_{\rm CF}=100$~mK, $T_{\rm QP} = 200$~mK, $\alpha=1.2\times 10^{-4}$, and $\Gamma_{\rm odd}=10$~MHz.
}
\label{fig:2dplot}
\end{centering}
\end{figure}

The measurement in figure~\ref{fig:2dplot} is made by stepping the microwave amplitude and sweeping the gate voltage with a 2.7 Hz triangular ramp. The data is sampled at 50 kHz and each ramp is sampled 8 times and averaged. The same measurement was repeated for several temperatures, figure~\ref{fig:temperature} shows selected
line traces of these measurements for different temperatures.

\section{Analysis of the results}\label{sec:results}

Figure~\ref{fig:2dplot} compares experimentally measured frequency shift to numerical results.
Two types of frequency-shift patterns appear, one at weak drive strengths and one at large
drive strengths,
with a pronounced change in the magnitude of the frequency-shift in between.
In this section, we make a detailed discussion of the processes
behind these features.
In simple terms, at low drive strengths we see effects due to interaction between the dressed charge-states
and the dissipative charge-fluctuator environment, and at large drive strengths it is the quasiparticle creation that
defines the frequency-shift pattern. We start by first analyzing the theoretical results due to interaction with each
of the environments alone, and after this analyze the situation when both of them are present, corresponding to the experiments.

\subsection{Characteristics in the presence of a charge-fluctuator environment}\label{sec:chargefluctuations}
\begin{figure}
\begin{centering}
\includegraphics[width=\linewidth]{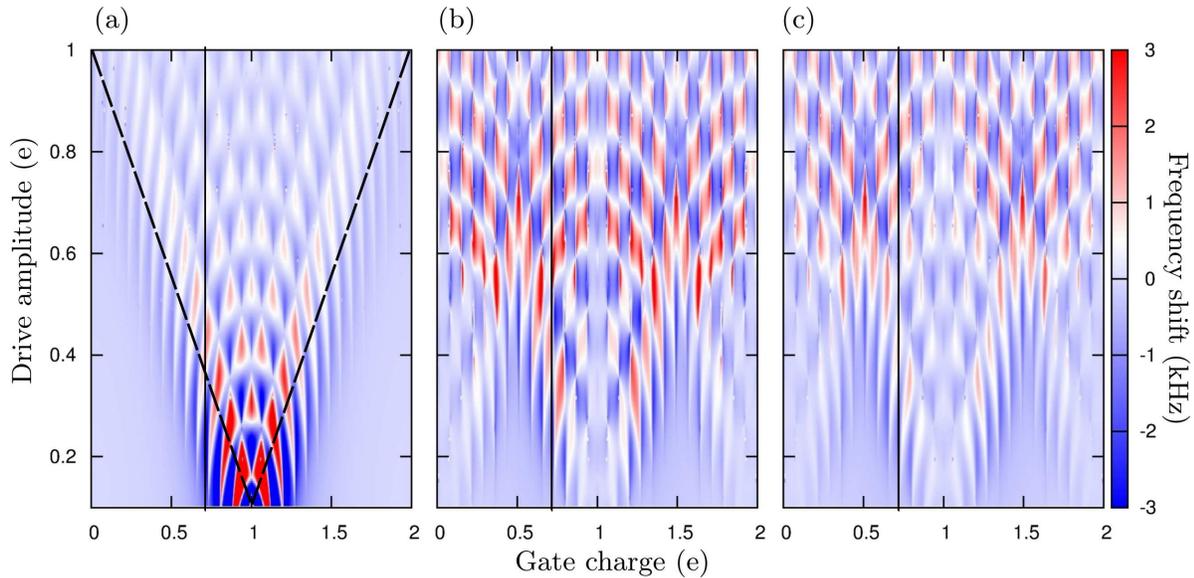}
\caption{(a) Theoretical frequency shift in the presence of only a charge-fluctuator environment.
A four-photon resonance is marked with a vertical line. At low drives, below the onset of the corresponding dressed-state coupling~(\ref{eq:CouplingAmplitudes}),
a bimodal behavior of the frequency shift occurs.
After the onset, triangle-shaped local minima/maxima regions appear.
The overall structure sketches an area that classically corresponds to gate-charge oscillations
that cover the degeneracy point $n_{\rm G}=e$ (dashed lines in figure~\ref{fig:2dplot}).
(b) In the presence of quasiparticles, bimodal behavior occurs at all drives due to a different type of system-environment coupling.
After the onset, rectangular-shaped local minima/maxima regions appear.
The characteristics change smoothly from $2e$ to $e$-periodic as the drive is increased.
(c) In the presence of charge fluctuators and quasiparticles, both of the previous features appear,
depending on which of the environments give the dominant contribution.
The results in (a) are restricted to the maximal value $\vert \langle \delta f_0 \rangle \vert=3$~kHz. Other parameters are given
in figure~\ref{fig:2dplot}.
}
\label{fig:numerics1}
\end{centering}
\end{figure}

In figure~\ref{fig:numerics1}a,  we plot numerical results for the frequency shift in the presence of the charge-fluctuator
environment (only).
This situation is considered also in Refs.~\cite{chris1,sill} and
the results discussed here are equivalent, but presented as the cavity frequency shift (that is the measured quantity here)
rather than the quantum capacitance.

As there is no quasiparticle tunneling included, we have $2e$-periodicity and symmetry
around the charging-energy degeneracy $n_{\rm G}=1$.
At low drive strengths a sign change is observed  when crossing the
the avoided level-crossings of dressed states, equation~(\ref{eq:ChargingEnergy}),
e.g.~at the $l=4$ resonance indicated by the vertical line in figure~\ref{fig:numerics1}.
At large drive strengths, a sequence of triangle-shaped local minimum and maximum regions appear.
The change between these two behaviors occurs roughly when
the semiclassical gate oscillations would cover the degeneracy $n_{\rm G}=1$, i.e.,
at the onset of the $l$:th order Bessel functions in amplitudes~(\ref{eq:CouplingAmplitudes}), indicated by the diagonal dashed lines in figure~\ref{fig:numerics1}a.

Below the onset, a bimodal structure appears due to population inversion at one side of the dressed-state resonances~\cite{chris2}. Here,
the dissipative transitions favor the dressed state with a higher amount of the lowest-energy charge-state.
This is because below the onset, the charge-state transition dominates, that needs the least amount of extra photons to be absorbed.
This population inversion can be understood using a two-charge states approximation (Appendix),
where the relevant dressed states (nearby the $l$:th order resonance, $ |l| >0 $) can be written as
$
\vert +;N\rangle=\sin (\phi/2)  \vert 0;N\rangle - \cos (\phi/2)  \vert 2; N-l\rangle
$, and
$\vert -; N\rangle= \cos(\phi/2)  \vert 0;N\rangle + \sin (\phi/2) \vert 2; N-l\rangle $.
Here
$\phi= \arctan\left(E^{\rm dr}_{\rm J} / \delta E^{\rm dr}_c \right)$
for $E_C^{\rm dr}>0$ and $\phi= \arctan\left( E^{\rm dr}_{\rm J} / \delta E^{\rm dr}_c \right)+\pi$ for $\delta E^{\rm dr}_C<0$.
We have $E_C^{\rm dr}=\delta E_C-l\hbar\omega_0$ and
$E_{\rm J}^{\rm dr}=E_{\rm J}J_{-l}(4A E_C/\hbar\omega_0)$.
If $\vert 0\rangle$ has a lower charging energy $E_C(\hat n-n_{\rm G})^2$, i.e., $n_{\rm G}<1$, and if the dressed-state transition rates
follow the concentrations of the state $\vert 0\rangle $,
then the populations are approximately $P_+=\sin^2(\phi/2)$,
and $P_-=1-P_+$.
Sweeping through the resonance (from left to right, $\delta E_C^{\rm dr}=+\infty\rightarrow -\infty$, $\phi= 0 \rightarrow \pi$),
switches the populations of the dressed charge states, and leads to the bimodal behavior.

Above the onset, the triangular form is a result of that
the coherent interaction between the charge states (\ref{eq:CouplingAmplitudes})
oscillate around zero with increasing drive.
The narrow part (and the simultaneous change of the sign) corresponds to the amplitude crossing zero,
whereas the wide part corresponds to a local maximum.
In this regime, dissipative processes with different photon-number changes are allowed, resulting in a normal (non-inverted) population of the dressed states.~\cite{chris2}

\subsection{Characteristics in the presence of quasiparticle environment}

\begin{figure}
\begin{centering}
\includegraphics[width=\linewidth]{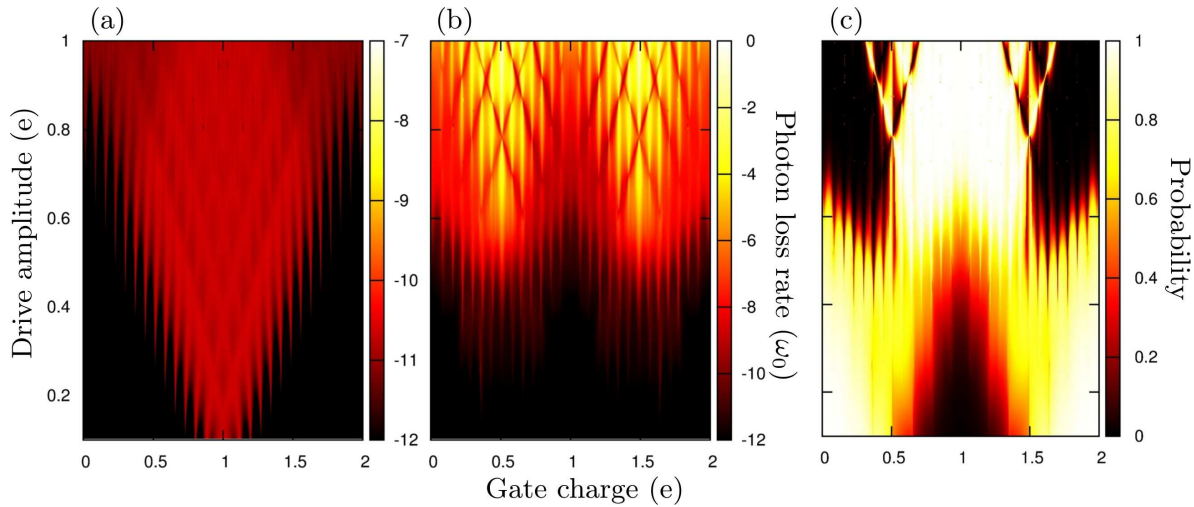}
\caption{Calculated photon loss rate, $\ln(\Gamma_{\rm Ph}/\omega_0$),
 (a) in figure~\ref{fig:numerics1}a (charge fluctuations) and (b) in figure~\ref{fig:numerics1}b (quasiparticles).
Note the different color scales.
For charge-fluctuator environment, (a), the photon dissipation rate follows the structure of dressed states, being most effective
for strong avoided level crossings of charge states. 
In the presence of thermal quasiparticles, (b),
the system avoids onsets of dressed-state resonances and dissipation stays low.
Escape to the other parity state occurs here through thermal excitation of the quasiparticles, which needs to overcome the charging energy difference.
At drive strengths that become comparable to the charging energy difference
the system also starts to populate the energetically higher parity state, and
the photon dissipation rate increases. For strong drive strengths quasiparticle creation appears with high dissipation rates.
(c) The theoretical probability that the system is in the even parity electron-number state in the simulation (b).
For low drives the system avoids (onsets of) dressed state resonances, since it wants to minimize the charging energy. At strong drives
quasiparticle creation favors the opposite parity, through establishing fast transitions from the parity space that was favored by the low drive
strength. The parameters in the simulations are the same as in figure~\ref{fig:2dplot}.
}
\label{fig:numerics2}
\end{centering}
\end{figure}

In figure~\ref{fig:numerics1}b,  we plot numerical results for the frequency shift in the presence of a quasiparticle
environment (only). When quasiparticles are introduced, both electron-number parities become relevant.
The two parities are not connected by the base Hamiltonian~(\ref{eq:hamiltonian1}),
but (perturbatively treated) quasiparticle tunneling leads to transitions between them.
The odd parity dressed states are copies of the even-parity ones, shifted by one
electron in the gate charge. One could then expect, that
the final frequency shift pattern is a weighted average of the results obtained
in section~\ref{sec:chargefluctuations}, with a relative shift of $e$ in the gate charge.
However, when compared to the case of charge fluctuations, the
populations of the dressed states inside a given parity are qualitatively different.
For example, for the parameters considered in this article,
quasiparticle tunneling always favors a bimodal structure around the $l$:th order resonance ( $\vert l \vert >0$), also above the onset
of the dressed-state resonances (following from the fact that photon-assisted quasiparticle tunneling is always well below its "onset").
The neighboring dressed-state resonances (neighboring $l$) have frequency shifts with opposite signs
[due to opposite derivatives of the amplitudes~(\ref{eq:CouplingAmplitudes})].
Combining this with the bimodality of dressed state populations at each resonance,
it follows that the frequency shifts of neighboring $l$ support each other,
and the total pattern consists of repeated rectangular-shaped regions of local minimum or maximum.

\subsubsection{Low drive strengths: Thermal and nonequilibrium quasiparticles}\label{sec:quasiparticles1}

At low drive strengths and at low temperatures the relative populations of the two electron-number parities also play an important role.
In the absence of thermal quasiparticles ($T_{\rm QP}=0$), the system prefers the even subspace,
and the island stays unpoisoned. 
This leads to a very strong frequency-shift near $n_{\rm G}=1$, similar as plotted in figure~\ref{fig:numerics1}a.
However, for quasiparticle temperatures observed in the
experiment ($T\sim 200$~mK), the average thermal quasiparticle density
is higher than the one of a single extra quasiparticle on the island, and
the system rather wants to minimize the charging energy. It then mostly populates the even subspace for $n_{\rm G}<1/2$ and the odd for $n_{\rm G}>1/2$, as seen in figure~\ref{fig:numerics2}c.
It follows, that the system avoids onsets of photon resonances and also the dissipation stays low, as seen in figure~\ref{fig:numerics2}b.
At the same time, the magnitude of the frequency shift is reduced, as is seen in figure~\ref{fig:numerics1}b
compared to figure~\ref{fig:numerics1}a.
The escape to the other parity state occurs here through thermal excitation of the quasiparticles, which needs to overcome the charging energy difference between the subspaces.
At drive strengths that become comparable to the charging-energy differences between the subspaces,
photon-assisted tunneling of thermalized quasiparticles takes place and the system also starts to populate the energetically higher parity state.
The photon dissipation rate increases, and at the dressed-state resonances ($\delta E_C^{\rm dr}=0$) such processes are enhanced.
Therefore, the corresponding increased dissipation and increased even-state population start with "spikes" in figures~\ref{fig:numerics2}b and~\ref{fig:numerics2}c.

\subsubsection{Large drive strengths: Quasiparticle creation}\label{sec:quasiparticles2}
The previous picture changes, when the quasiparticle tunneling
with Cooper-pair breaking becomes the dominant quasiparticle process.
In the corresponding frequency shift, figure~\ref{fig:numerics1}b, this corresponds to the start of the enhanced
frequency-shift region ($A\sim 0.6$),
and in figure~\ref{fig:numerics2}b to
the first drastic increase in dissipation (beginning of the
bright red region).
Away from charge degeneracies $n_{\rm G}=0.5$ and $n_{\rm G}=1.5$,
the parity of the system gets strongly polarized, see figure~\ref{fig:numerics2}c,
since tunneling out of one parity space dominates.
For a substantial density of thermal quasiparticles
(so that the single extra unpaired electron on the island loses its importance), as in figure~\ref{fig:numerics2}b,
the favored parity  is the opposite of the low-drive case.
When the drive is increased further, transitions out of the favoured parity state gets stronger (beginning
of the yellow region in figure~\ref{fig:numerics2}b).

Such successive inter-parity transitions with quasiparticle creation can be understood in the two-charge-state approximation (Appendix).
For an explicit discussion, we now assume that we have $n_{\rm G}\sim 3/4$,
where the two relevant even-parity charge states are $\vert 0\rangle$ and $\vert 2\rangle$, and the odd-parity charge states
$\vert 1\rangle$ and $\vert -1\rangle$.
These are connected via coherent Cooper-pair tunneling and exchange of $l_e$ (even) or $l_o$ (odd) photons with the cavity.

\begin{figure}
\begin{centering}
\includegraphics[width=0.9\linewidth]{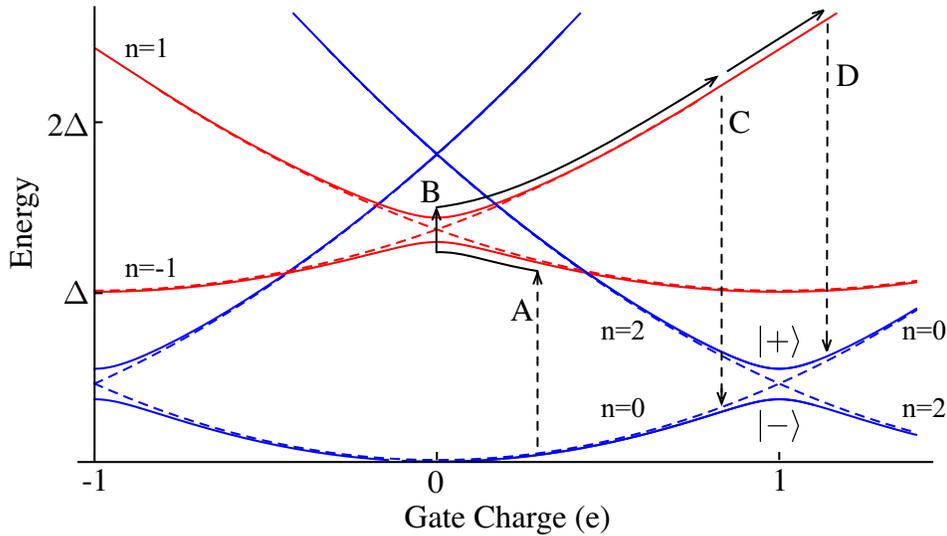}
\caption{
A semiclassical illustration of quasiparticle mediated population inversion and simultaneous parity recovery.
A: The system is at some point brought to the odd subspace, for example,
by a tunneling of a nonequilibrium quasiparticle.
B: If the drive is strong enough, the classical driving field will bring the system to the odd degeneracy point ($n_{\rm G}=0$), where coherent tunneling of a Cooper pair will evolve the system into the excited band. C and D : The driving field brings the system to energies allowing quasiparticle
tunneling ($E> 2\Delta$), and this will rapidly cause a pair-breaking event with a transition back to the even subspace.
This transition from $\vert -1\rangle$ favors the charge state $\vert 0\rangle$, since relaxation to $\vert 2\rangle$ would require a simultaneous Cooper-pair tunnelling. This results in an inversion of the population of the
dressed states $|-\rangle$ and $|+\rangle$,
on either side of the degeneracy point, as
the charge concentrations of $|-\rangle$ and $|+\rangle$ are highly dependent on charge bias.
We visualize this dependence as two different places for the relaxation process, C and D, that favor different dressed states.
}
\label{fig:visualization}
\end{centering}
\end{figure}

The following picture emerges: when being in the charge state $\vert 0 \rangle$, coherent Cooper-pair tunneling takes
the system to the more energetic state $\vert 2\rangle$, through the help of $l_e$ cavity photons.
Occasionally transitions from the even to the odd subspace occurs,
mostly through the channel
$\vert 2\rangle\rightarrow \vert 1\rangle$ and rate $\Gamma_{2\rightarrow 1}^{\rm QP}$. This transition channel
dominates over $\Gamma_{0\rightarrow 1}^{\rm QP}$, as it requires
the least amount of extra photons $M_e$ ($l_e+M_e\sim 2\Delta/\hbar\omega_0\sim 14$).
This is not a fast process, due to a
rather large $M_e$. When being in the state $\vert 1\rangle$, coherent Cooper-pair tunneling couples the system to the
state $\vert -1\rangle$, which has the highest charging energy, and requires $l_o>l_e$ photons. However,
the quasiparticle transition $\vert -1\rangle \rightarrow \vert 0\rangle$
needs now a relatively low amount of extra photons, $M_o<M_e$, and
its rate $\Gamma_{-1\rightarrow 0}^{\rm QP}$ is therefore the largest in this interplay.
The system thus switches quickly back to the even-parity space.
This interplay is visualized in figure~\ref{fig:visualization}, in a simplified semiclassical picture.

The even-parity dressed state, that this process favors, is the state with most concentration of the even-parity charge state $\vert 0\rangle$,
as this is the only state that can be directly accessed
from the energetic state $\vert -1 \rangle$ through quasiparticle tunneling. Similarly as for the case of charge fluctuations, this can either be a ground or an excited dressed-state, depending on which side of the resonance condition we are biased, see the discussion in section~\ref{sec:chargefluctuations}. Therefore, the enhanced frequency shift pattern in figure~\ref{fig:numerics1}b, at the large drive strengths, is both due to the bimodal behavior of the dressed-state populations,
and because the system practically populates only this sensitive parity subspace, see the even-parity populations in figure~\ref{fig:numerics2}c.
The observed frequency-shift pattern can be derived analytically in this region, see Appendix.

\subsection{Both environments present simultaneously}
In figures~\ref{fig:2dplot} and~\ref{fig:numerics1}c, we plot numerical results for the frequency-shift pattern in the presence of both
environments, charge fluctuations and quasiparticles.
The central result is that there exists regions (in amplitude $A$) where each environment is dominating:
for low $A$ charge fluctuations dominate and for high $A$ the quasiparticle generation rate overcomes
the charge fluctuator decoherence.

In numerical simulations that fit to the experiment,
we use a relatively high quasiparticle temperature ($T_{\rm QP}=200$~mK), which leads to a reduced frequency shift
at low drive strengths compared to the quasiparticle-free situation, see figure~\ref{fig:numerics1}a. This is because the system spends
most of its time in a parity state that does not produce any
coherent dynamics (section~\ref{sec:quasiparticles1}).
However, at low drives the intra-parity rates due to the charge fluctuations overcome the parity-switching rates due to quasiparticles.
Therefore, it is the charge fluctuations which define the
intra-parity populations and the observed interference pattern (the triangular form, seen also in figure~\ref{fig:numerics1}a).
The triangular form of local minima or maxima is qualitatively different from the high-drive region, where a rectangular-shaped frequency-shift pattern is observed. Also the number of minima and maxima on a horizontal line changes at the cross-over.
A cross-over occurs when the quasiparticle tunneling rate out of the subspace, in which the system spends most of its time,
overcomes the intra-parity transition rates due to charge fluctuations, which want to normalise the dressed-state populations.
For large drive strengths, it is therefore the quasiparticle creation in both of the parity subspaces that dominates 
the transition rates, and the relevant physics is the same as described in section~\ref{sec:quasiparticles2}.

\begin{figure}
\begin{centering}
\includegraphics[width=0.9\linewidth]{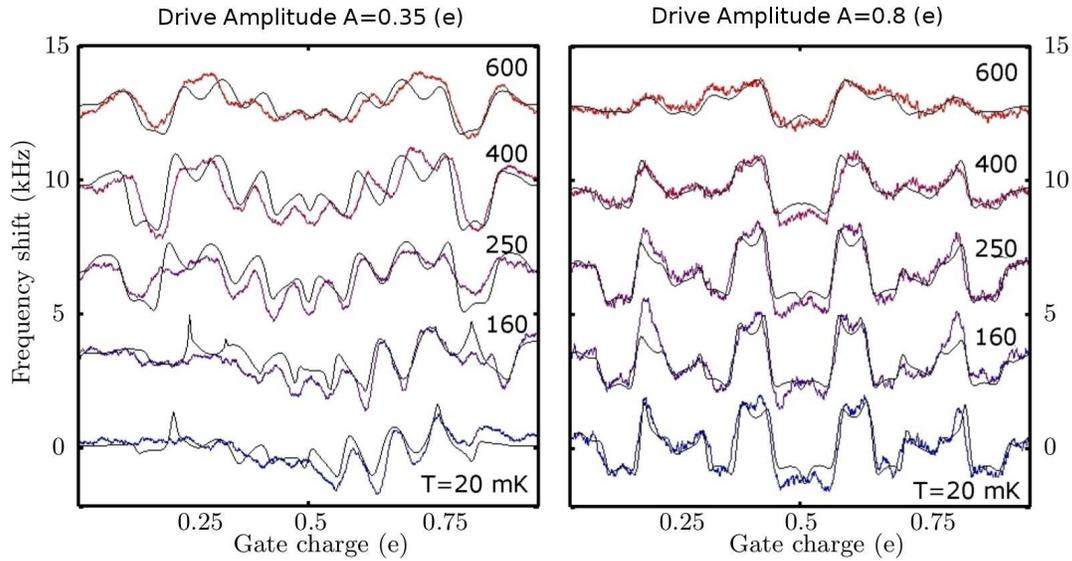}
\caption{
Left: Measured frequency shift (colored line) compared with the numerical results (black line) as a function of gate charge for a constant microwave drive strength $A=0.35$ and for
several temperatures (shifted by 3~kHz). Right: The same data for the drive amplitude $A=0.8$, where the frequency shift has only weak temperature dependence.
This frequency shift pattern is due to the bimodal behavior of the dressed state populations due to quasiparticle tunneling
with Cooper-pair breaking, discussed in section~\ref{sec:quasiparticles2}.
In the numerical model the lowest environmental temperatures were set to $T_{\rm CF}=100$~mK and $T_{\rm QP} = 200$~mK, otherwise  they were
changed according to given temperatures.
Additional Gaussian gate-charge broadening $\sigma_n=0.0052e$, $0.0057e$, $0.0064e$, $0.0095e$, $0.0141e$
was applied to the numerical results at the different temperatures,
to account for increasing low-frequency gate-charge fluctuations.
This amount of extra broadening is consistent with other works~\cite{TempDep1,TempDep2}.
Otherwise the parameters are the same as in figure~\ref{fig:2dplot}.
}
\label{fig:temperature}
\end{centering}
\end{figure}

\subsection{Temperature dependence}
Line traces of the temperature dependence at two drive
strengths are shown in figure~\ref{fig:temperature}. The most important observation here is that
for the low drive strength ($A=0.35$) there is a noticeable change in the form and the periodicity in the frequency-shift pattern
when the temperature is increased,
whereas for the large drive amplitude ($A=0.8$) the data is very robust under the change of temperature.
For the low drive strength the periodicity changes from $2e$ to $e$ around 250~mK,
where the parities become symmetrically populated due to a large number of thermal quasiparticles.
Also in the region $0.8<n_{\rm G}<1$ a new sensitive pattern starts to appear with increasing temperature.
This since at higher temperatures the sensitive parity space,
that produces an observable frequency shift, becomes more populated.
For even higher temperatures, the small structure at $0.3<n_{\rm G}<0.7$ tends to vanish,
as the temperature becomes comparable to the dressed energy-level splittings.

At the drive amplitude $A=0.8$ the $e$-periodic frequency-shift pattern is due to quasiparticle tunneling with Cooper-pair breaking,
described in section~\ref{sec:quasiparticles2}. Raising the temperature, equivalent to raising the quasiparticle density,
does not influence the observed structure until temperatures as high as $T\sim 0.4$~K. Before this, there exists more and more
quasiparticles, but the rate induced by their presence stays below the quasiparticle creation rate.
Even though a thermal or nonequilibrium quasiparticle takes the system to the other parity state, it is brought
back quickly via the quasiparticle creation process.
The rate for quasiparticle creation starts to change significantly
only for very high temperatures, $k_{\rm B}T\sim \Delta/2$.
Finally, the recombination rate (not considered here)
is fast enough to keep the
nonequilibrium quasiparticle density moderate.
{\em Thus, the presence of nonequilibrium
quasiparticles does not influence the dynamics at high drives}.

\section{Conclusions and outlook}\label{sec:Discussion}
In this work, we studied a circuit-QED realization of a strongly driven artificial atom coupled to a microwave cavity.
We experimentally found a new regime at very large drive strengths, where the measured system properties showed increased
sensitivity to the qubit gate charge and also robustness against environmental decoherence.
To explain the observed effects, we extended the dressed-state description of this system
to account for the effect of quasiparticle tunneling at the Josephson junction.
We found that the new properties are due to quasiparticle creation processes,
that involve exchange of a large number of photons between the cavity and the artificial atom.
This process is dominant even at elevated temperatures, since the
quasiparticle creation rate becomes higher than other environmental
processes in the strong driving regime.
The observed interference pattern is robust with respect to quasiparticle poisoning, temperature, and drive strength.
Using the developed theoretical methods we can extract many properties of the driven
cavity-atom system, properties related to the environment, and also properties related to quasiparticle
dynamics. Within the discussed regime, the system naturally lends
itself to being used as a very sensitive and robust charge detector.

\ack
We acknowledge the Swedish Research Council VR, EU FP7 programme under the grant agreement 'ELFOS', the Marie Curie Initial Training Action (ITN) Q-NET 264034, the NMS pathfinder program, and the Linnaeus centre for quantum engineering for financial support.

\appendix

\section*{Appendix: Two-level approximation nearby dressed-state resonances}
\subsection*{Hamiltonians of the two parities}
As a special case, we find a two-level system in the odd subspace, for example, when $\hat n$ is restricted to the values $\pm 1$ instead of $0$ and $2$.
We have then two blocks in the total Hamiltonian, corresponding to the two
parities,
\begin{eqnarray}
H^{\rm even}&=&\hbar\omega_0 a^{\dagger}a+\frac{\delta E_C^{\rm even}}{2}\sigma_z  -\frac{E_J}{2}\sigma_x+g(a^{\dagger}+a)\sigma_z \\
H^{\rm odd}&=&\hbar\omega_0 a^{\dagger}a+\frac{\delta E_C^{\rm odd}}{2}\sigma_z + E_C(2n_{\rm G}-1) -\frac{E_J}{2}\sigma_x+g(a^{\dagger}+a)\sigma_z.
\end{eqnarray}
We assume now $-1<n_{\rm G}<1$. The
charging-energy difference in the even Hamiltonian is as before,
$\delta E_C^{\rm even}=4E_C(1-n_{\rm G})$,
but for the odd one we get
$\delta E_C^{\rm odd}=4E_Cn_{\rm G}$.
Otherwise the Hamiltonian parameters are the same as before.
A convenient basis is the displaced number states,
\begin{equation}\label{eq:basis2level}
\vert \uparrow/\downarrow;N\rangle=\exp[\mp\beta( a^{\dagger}- a)]\vert\uparrow/\downarrow\rangle\vert N\rangle= D(\mp\beta)\vert\uparrow/\downarrow\rangle\vert N\rangle.
\end{equation}
These are eigenstates of Hamiltonian (\ref{eq:hamiltonian1}) for $E_{\rm J}=0$ and $\beta=g/\hbar\omega_0$,
with the corresponding eigenenergies
\begin{equation}\label{eq:dressed_energies}
E_{\uparrow/\downarrow N}=N\hbar\omega_0-\frac{ g^2}{\hbar\omega_0}\pm \frac{\delta E_C}{2}.
\end{equation}
In the considered semiclassical limit, $N\gg 1$, the corresponding matrix elements are
$\langle \uparrow;N+l\vert \sigma_x \vert \downarrow;N\rangle \approx J_{l}\left( 4\beta\sqrt{N}\right)$, and
$\langle \downarrow;N+l\vert \sigma_x \vert \uparrow;N\rangle \approx J_{l}\left( -4\beta\sqrt{N}\right)$.

\subsection*{Eigenstates}
Near the $l$-photon resonance  we can reduce the description
of the system to the dressed two-charge-state Hamiltonian~\cite{review},
\begin{equation}
H_{\rm dr}=\frac{\delta E_{C}^{\rm dr}}{2}\sigma_z - \frac{E^{\rm dr}_{\rm J}}{2}\sigma_x.
\end{equation}
Here $\delta E^{\rm dr}_C=\delta E_C-l\hbar\omega_0$ and $E^{\rm dr}_{\rm J}=E_{\rm J} J_{-l}\left( 4E_C A/\hbar\omega_0\right)$.
The Pauli $\sigma$-matrices operate now in the Hilbert space spanned by $\vert \downarrow;N\rangle$
and $\vert \uparrow;N-l\rangle$.
This gives the eigenstates
\begin{eqnarray}
\vert +;N\rangle=\sin \frac{\phi}{2}  \vert \downarrow;N\rangle - \cos \frac{\phi}{2}  \vert \uparrow; N-l\rangle \label{eq:eigenstates1}, \\
\vert -; N\rangle= \cos\frac{\phi}{2}  \vert \downarrow;N\rangle + \sin \frac{\phi}{2} \vert \uparrow; N-l\rangle \label{eq:eigenstates2},
\end{eqnarray}
with the corresponding eigenenergies $E_{\pm}=\pm\frac{1}{2}\sqrt{ ( \delta E^{\rm dr}_C )^2+( E^{\rm dr}_{\rm J} )^2}$.
Here $\phi= \arctan\left(E^{\rm dr}_{\rm J} / \delta E^{\rm dr}_c \right)$
for $\delta E^{\rm dr}_C>0$ and $\phi= \arctan\left( E^{\rm dr}_{\rm J} / \delta E^{\rm dr}_c \right)+\pi$ for $\delta E^{\rm dr}_C<0$. It is important to notice that the level splitting
is a function of $N$, leading to oscillations as a function of drive amplitude $A$. This is the dressed state description of the
Landau-Zener-St\"uckelberg interference effect of a periodically driven two-level system~\cite{chris2,chris1,sill,review}.

\subsection*{Frequency shift analytically}\label{sec:analytical1}
Using the idea of section~\ref{sec:FreqShift}, and the two-level approximation nearby the $l$-photon resonance,
we obtain
\begin{equation}\label{eq:ShiftCF}
\delta f_0^{\pm} \approx \frac{2 E_{\rm J} E_{\rm J}^{\rm dr}(a)}{a h}\left(\frac{G C_{\rm G}}{C_{\rm J}} \right)^2  \frac{ J_{l-1}(a)  - J_{l+1}(a)  }{\sqrt{[\delta E_C^{\rm dr}(a)]^2+ [E_{\rm J}^{\rm dr}(a)]^2}}.
\end{equation} 
Here $a=4\beta\sqrt{N}$.
This implies a maximal frequency shift of the order $(E_{\rm J}/h)G^2(C_{\rm G}/C_{\rm J})^2$, for
the experiment considered here meaning $\sim 100$~kHz (for a $10$~GHz oscillator). However,
the measured shift is of the kHz-scale,
which we find to be due to quasiparticle poisoning effects
and due to partial cancellation of the two opposite shifts,
discussed in more detail in section~\ref{sec:results}.
The dissipative shift of the oscillator, due to its finite $Q$-factor, is of the order of a few Hz,
which is not observed here.

In the presence of only charge fluctuators (no quasiparticles) and below the onset of $l$:th dressed state gap ($l$:th order Bessel function),
the populations of the eigenstates show bimodal behavior, favoring the state of a lower charging energy.
The resulting analytical form of the frequency shift is then similar
as calculated below for the case of quasiparticle dominated region.

\subsection*{Frequency-shift pattern in the quasiparticle creation dominated region}
The high polarization of the parity at high drives, figure~\ref{fig:numerics2}c, makes it possible
to derive analytical expressions for the observed frequency-shift pattern. We assume now that the quasiparticle tunneling rates stay below the energy-level splittings of the dressed states (a more detailed condition is derived below).
The rates  are practically incoherent summations of contributions
coming from transitions between different charge states.
We assume now $n_{\rm G}\sim 3/4$. Here, the dominating contribution to the transition rate
between the dressed charge states is
\begin{equation}
\Gamma^{\rm QP}_{-1\rightarrow 0}\sim \frac{\Delta\vert J_{M}(2\beta\sqrt{N})\vert^2}{e^2 R_{\rm T}} \, , \,\,\,\, M\approx \frac{2\Delta-3E_C}{\hbar\omega_0}.
\end{equation}
The bottleneck rate is
\begin{equation}
\Gamma^{\rm QP}_{2 \rightarrow 1}\sim \frac{\Delta\vert J_{M'}( 2\beta\sqrt{N})\vert^2}{e^2R_{\rm T}} \, , \,\,\,\, M' \approx \frac{2\Delta-3E_C/2}{\hbar\omega_0}>M.
\end{equation}

The transitions from the even subspace dressed state $\vert +\rangle_e$ occur predominantly via the charge state $\vert 2\rangle$ to
odd charge state $\vert 1\rangle$,
meaning that the rate between the dressed states is proportional to $\cos^2\frac{\phi}{2}$. Similarly, the fast transition back occurs
via the charge state $\vert -1\rangle$ to the charge state $\vert 0\rangle$.
The overall rate from the even-parity state $\vert +\rangle_e $ to the even-parity state $\vert -\rangle_e $ becomes then
proportional to $\cos^4\frac{\phi}{2}$. Similarly, the rate from $\vert -\rangle_e $ to $\vert +\rangle_e $ is proportional to $\sin^4\frac{\phi}{2}$. We mark these two effective rates by
$\Gamma_{+-}$ and $\Gamma_{-+}$, respectively.
By this argumentation we get for the populations inside the even-parity subspace (populations in the odd space are close to zero)
\begin{equation}
P_{\pm}=\frac{\Gamma_{\mp\pm}}{\Gamma_{+-}+\Gamma_{-+}}=\frac{1}{2}\left( 1\pm\frac{2\cos\phi}{1+\cos^2\phi} \right)
\end{equation}

The small photon-energy shift when jumping between nearby equivalent (photon-translated) states
was estimated in~(\ref{eq:ShiftCF}). This can be now expressed as
\begin{equation}
h\delta f_0^+\approx  \sin\phi J'_l({ a })\frac{4E_{\rm J}}{{ a}} \left(\frac{G C_{\rm G}}{C_{\rm J}} \right)^2 .
\end{equation}
Here $\vert l \vert >0$,
$\sin\phi=E_{\rm J}^{\rm dr}/\sqrt{(E_{\rm J}^{\rm dr})^2+(\delta E_C^{\rm dr})^2}$, and
$\cos\phi=\delta E_C^{\rm dr}/\sqrt{(E_{\rm J}^{\rm dr})^2+(\delta E_C^{\rm dr})^2}$. The average shift has the form
\begin{equation}
h\delta f_0^+(P_+-P_-)=\frac{2\sin \phi\cos\phi}{1+\cos^2\phi} J'_l({ a})\frac{4E_{\rm J}\beta^2}{{ a}}.
\end{equation}
As a function of the qubit parameters this has the form
\begin{eqnarray}\label{eq:freqshift}
\langle \delta f_0 \rangle=\frac{8\beta^2}{{ha}}\frac{\delta E_C^{\rm dr} E^2_{\rm J} J_l({ a})}{2(\delta E_C^{\rm dr})^2 +[E_{\rm J} J_l({ a})]^2}J'_l({ a}).
\end{eqnarray}
According to our simulation the observed frequency shift follows this approximation well.
The charge sensitivity is proportional to derivative with respect to the $\delta E_C^{\rm dr}$.
The maximum value of this ($\delta E_C^{\rm dr}\rightarrow 0$) becomes then
 $4\beta^2 \left[ J_{l-1}({ a}) - J_{l+1}({ a}) \right]/{ ha}J_l({ a})$. It is independent of $E_{\rm J}$.
It diverges when the $l$:th order Bessel function goes to zero. Around this special point the frequency shift
also changes sign. The divergence is an artefact from approximating the frequency change with a derivative.
In our simulations the divergence is also damped out as the incoherent transition rates exceed the dressed state splitting
nearby the point where the dressed state splitting goes to zero.

\subsection*{Validity region of the analysis for the quasiparticle creation regime}
Let us now consider when the discussed interpretation of the quasiparticle creation dominated region is valid.
First, in order to keep ourselves in a simple picture of two relevant charge states in each parity subspace, we require that the drive
stays weaker than the charging energy of Cooper pairs. This means that we restrict to the drive region
\begin{equation}\label{eq:condition1}
A < 1.
\end{equation}
Another demand is that the quasiparticle tunneling rates stay below
the dressed-state splittings. We consider here the region $n_{\rm G}\approx 1/4+m/2$,
where $m$ is an integer.
The relevant charging-energy differences in this region are roughly $\sim 2E_C$.
The number of photons $l'$
needed from the oscillator to assist the tunneling is defined as $l' \sim 2(\Delta-E_C)/\hbar\omega_0$.
On the other hand, $l$ photons are exchanged in the resonant interplay to form the dressed-states splitting
$\Delta_l=E_{\rm J}\vert J_{l}(4\beta) \vert $, where $l\sim 2E_C/\hbar\omega_0$ and
$l+l'\sim 2\Delta/\hbar\omega_0$.
Comparison between the dressed-state splitting
and the photon-assisted tunneling rate leads to a demand
$
\frac{E_{\rm J}}{\hbar} \vert J_{l}(4\beta\sqrt{N})\vert \gg \frac{2\Delta}{e^2R_{\rm T}} \vert J_{l'}(2\beta\sqrt{N}) \vert^2
$.
Here we can use the Ambegaokar-Baratoff formula $R_{\rm T}=R_{\rm Q}\Delta/2E_{\rm J}$ ($T=0$) and obtain
\begin{equation}\label{eq:condition3}
\vert J_{l}(4\beta\sqrt{N})\vert \gg  \vert J_{l'}(2\beta\sqrt{N}) \vert^2.
\end{equation}
We want to operate in a region where the left-hand side is above its threshold, but the right-hand side
is still well below it. As the number $l$ is defined by $E_C$ alone, $l'$ depends on $E_C$ and $\Delta$.
For too low $\Delta$, condition~(\ref{eq:condition3}) is always violated.

\bibliographystyle{plainnat}

\begin{thebibliography}{26}

\bibitem{GeneralCircuitQED} Devoret M H and Schoelkopf R J 2013 {\it Science} {\bf{339}} 1169 

\bibitem{Decoherence2005} Ithier G, Collin E, Meeson P J, Vion D, Esteve D, Chiarello F, Shnirman A, Makhlin Y, Schriefl J  and Sch\"on G 2005 {\it Phys. Rev.} B {\bf{72}} 134519

\bibitem{BathEngineeringPRL1996} Poyatos J F, Cirac J I and Zoller P 1996 {\it Phys. Rev. Lett.} {\bf{77}} 4728

\bibitem{chris2} Wilson C M, Johansson G, Duty T, Persson F, Sandberg M and Delsing P 2010 {\it Phys. Rev.} B {\bf{81}} 024520

\bibitem{chris1} Wilson C M, Duty T, Persson F, Sandberg M, Johansson G and Delsing P 2007 {\it Phys. Rev. Lett.} {\bf{98}} 257003

\bibitem{sill} Sillanp\"a\"a M, Lehtinen T, Paila A, Makhlin Y and Hakonen P 2006 {\it Phys. Rev. Lett.} {\bf{96}} 187002

\bibitem{review} Shevchenko S N, Ashhab S and Nori F 2010 {\it Phys. Rep.} {\bf{492}} 1

\bibitem{Nakamura} Nakamura Y, Pashkin Y A and Tsai J S 1999 {\it Nature} {\bf{398}} 786

\bibitem{black} Black E D 2001 {\it Am. J. Phys.} {\bf{69}} 79

\bibitem{tobias} Lindstr\"om T, Burnett J, Oxborrow M and Tzalenchuk A Y 2011 {\it Rev. Sci. Inst.} {\bf{82}} 104706

\bibitem{ShortPaper} de Graaf S E, Lepp\"akangas J, Adamyan A, Danilov A D, Lindstr\"om T, Fogelstr\"om M, Bauch T, Johansson G and Kubatkin S E 2013 arXiv:1306.4465

\bibitem{LutchynFirst} Lutchyn R, Glazman L and Larkin A 2005 {\it Phys. Rev.} B {\bf{72}} 014517

\bibitem{Aumentado} Aumentado J, Keller M W, Martinis J M and Devoret M H 2004 {\it Phys. Rev. Lett.}  {\bf{92}} 066802

\bibitem{Lenander} Lenander M, Wang H, Bialczak R C, Lucero E, Mariantoni M, Neeley M, O'Connell A D, Sank D, Weides M, Wenner J, Yamamoto T, Yin Y, Zhao J, Cleland A N and Martinis J M 2011 {\it Phys. Rev.} B {\bf{84}} 024501

\bibitem{Sun} Sun L, DiCarlo L, Reed M D, Catelani G, Bishop L S, Schuster D I, Johnson B R, Yang G A, Frunzio L, Glazman L, Devoret M H and Schoelkopf R J 2012 {\it Phys. Rev. Lett.}  {\bf{108}} 230509


\bibitem{juha1} Marthaler M, Lepp\"akangas J and Cole J H 2011 {\it Phys. Rev.} B {\bf{83}} 180505(R)

\bibitem{Metamaterials} Fazio R and van der Zant H 2001 {\it Phys. Rep.} {\bf 355} 235

\bibitem{Metamaterials2} Goldstein M, Devoret M H, Houzet M and Glazman L I 2013 {\it Phys. Rev. Lett.} {\bf 110} 017002

\bibitem{QOBook} Walls D F and Milburn G J 2008 {\it Quantum Optics} (Berlin: Springer)

\bibitem{Weiss} Weiss U 1999 {\it Quantum Dissipative Systems} 2nd ed. (Singapore: World Scientific)

\bibitem{Astafiev} Astafiev O, Pashkin Y A, Nakamura Y, Yamamoto T and Tsai J S 2004 {\it Phys. Rev. Lett.} {\bf{93}} 267007

\bibitem{Bloch} Bloch F 1957 {\it Phys. Rev.} {\bf{105}} 1206

\bibitem{Redfield} Redfield A G 1957 {\it IBM J. Res. Dev.} {\bf{1}} 19

\bibitem{leggett} Leggett A J, Chakravarty S, Dorsey A T, Fisher M P A, Garg A and Zwerger W 1987 {\it Rev. Mod. Phys.} {\bf 59} 1

\bibitem{Tinkham} Tinkham M 1996 {\it Introduction to superconductivity} 2nd ed. (New York: McGraw-Hill).

\bibitem{juha3} Lepp\"akangas J and Thuneberg E 2008 {\it Phys. Rev.} B {\bf{78}} 144518

\bibitem{Lutchyn} Shaw M D, Lutchyn R M, Delsing P and Echternach P M 2008 {\it Phys. Rev.} B {\bf{78}} 024503

\bibitem{Martinis} Martinis J M, Ansmann M and Aumentado J 2009 {\it Phys. Rev. Lett.} {\bf{103}} 097002

\bibitem{Catelani} Catelani G, Koch J, Frunzio L, Schoelkopf R J, Devoret M H and Glazman L I  2011 {\it Phys. Rev. Lett.} {\bf{106}} 077002

\bibitem{juha4} Lepp\"akangas J and Marthaler M 2012 {\it Phys. Rev.} B {\bf{85}} 144503

\bibitem{Schon} Sch\"on G and Zaikin A D 1994 {\it Eur. Phys. Lett.} {\bf{26}} 695

\bibitem{juha5} Lepp\"akangas J, Marthaler M and Sch\"on G 2011 {\it Phys. Rev.} B {\bf{84}} 060505(R)

\bibitem{shnirman} Shnirman A and Makhlin Y 2003 {\it JETP Lett.} {\bf{78}} 447








\bibitem{fractalpaper}de Graaf S E, Danilov A V, Adamyan A, Bauch T and Kubatkin S E 2012 {\it J. Appl. Phys.} {\bf{112}} 123905

\bibitem{kim2012}Kim Z, Suri B, Zaretskey V, Novikov S, Osborn K D, Mizel A, Wellstood F C and Palmer B S 2012 {\it Phys. Rev. Lett.} {\bf{106}} 120501

\bibitem{houck2008} Houck A A, Schreier J A, Johnson B R, Chow J M, Koch J, Gambetta J M, Schuster D I, Frunzio L, Devoret M H, Girvin S M and Schoelkopf R J 2008 {\it Phys. Rev. Lett.} {\bf{101}} 080502


\bibitem{jonspaper} Burnett J, Lindstr\"om T, Oxborrow M, Harada Y, Sekine Y, Meeson P and Tzalenchuk Y A 2013 {\it Phys. Rev.} B {\bf{87}} 140501

\bibitem{oneoverfnoise} Dutta P and Horn P M 1981 {\it Rev. Mod. Phys.} {\bf{53}} 497

\bibitem{degraaf2013} de Graaf S E, Danilov A V, Adamyan A and Kubatkin S E 2013 {\it Rev. Sci. Instr.} {\bf{84}} 023706

\bibitem{TempDep1} Astafiev O, Pashkin Yu A, Nakamura Y, Yamamoto T and Tsai J S 2006 {\it Phys. Rev. Lett.} {\bf{96}} 137001

\bibitem{TempDep2} Gustafsson M V, Pourkabirian A, Johansson G, Clarke J and Delsing P 2012 arXiv:1202.5350




\end{thebibliography}

\end{document}